\documentclass[10pt,aps,prc,showpacs,twocolumn,nofootinbib]{revtex4}
\usepackage[english]{babel}
\usepackage{graphicx}
\usepackage{epsfig}
\usepackage{amsmath}
\usepackage{amssymb}
\usepackage{latexsym}
\usepackage{float}

\usepackage[usenames]{color}

\begin{document}
\title{Uniform description of bulk observables in the hydrokinetic model of A+A collisions at RHIC and LHC}

\author{Iu.A. Karpenko$^{1,2}$}
\author{Yu.M. Sinyukov$^{1}$}
\affiliation{$^{1}$Bogolyubov Institute for Theoretical Physics,
Metrolohichna str. 14b, 03680 Kiev,  Ukraine}
\affiliation{$^{2}$Frankfurt Institute for Advanced Studies,
Ruth-Moufang-Str. 1, 60438 Frankfurt am Main, Germany}
\author{K. Werner}
\affiliation{SUBATECH, University of Nantes -- IN2P3/CNRS -- EMN, Nantes, France}

\begin{abstract}
A simultaneous  description  of hadronic yields;  pion, kaon, and proton spectra; elliptic flows; and femtoscopy scales in the hydrokinetic model of A+A collisions is presented at  different centralities for the top BNL Relativistic Heavy Ion Collider (RHIC) and CERN Large Hadron Collider (LHC) 2.76-TeV  energies. The initial conditions are based on the Glauber Monte-Carlo simulations. When going from RHIC to LHC energy in the model, the only parameters changed are the normalization of the initial entropy defined by the number of all charged  particles in most central collisions, contribution to entropy from binary collisions and barionic chemical potential. The hydrokinetic model is used in its hybrid version (hHKM), which provides the correct match (at the isochronic hypersurface) of the decaying hadron matter evolution with hadronic ultrarelativistic quantum molecular dynamics cascade. The results  are compared  with the standard hybrid models where  hydrodynamics and hadronic cascade are matching just at the non-space-like hypersurface of chemical freeze-out or on the isochronic hypersurface. The modification of the particle number ratios at LHC caused, in particular, by the particle annihilations at the afterburn stage is also analyzed.
\end{abstract}

\pacs{13.85.Hd, 25.75.Gz}
 \maketitle

\section{Introduction}
Soon after the first Large Hadron Collider (LHC) heavy ion results were launched,  it became evident that the hydrodynamic  picture of the collision processes, confirmed at Relativistic Heavy Ion Collider (RHIC),, is  clearly seen also at much higher energy. This conclusion is based on two classes of observables. The first one is related to the azimuthal anisotropy of particle  spectra expressed basically through its second harmonics, or $v_2$ coefficients. The obtained LHC results for the transverse momentum dependence of $v_2(p_T)$ at a given centrality bin were found to be similar to the ones at RHIC, except for the higher momentum range at LHC  \cite{ALICE_0}. This is the evidence of the same hydrodynamic mechanism of the anisotropy formation  as at RHIC.   The second type of observables deals with the direct measurements of the space-time structure of nucleus-nucleus collisions by means of the correlation femtoscopy. The femtoscopic spatiotemporal scales of the systems are typically represented in terms of the interferometry radii. The hydrodynamic predictions \cite{LHC predictions} for $p_T$ behavior of the radii at the LHC energies were confirmed by the ALICE experiment \cite{aliceHBT}. The most impressive hydrodynamic prediction \cite{Karp-Sin, Cracow_0}, that the ratio of the two transverse interferometry radii, {\it out} to {\it side}, will drop in the whole $p_T$- interval with increasing collision energy and reaches a value close to unity at the LHC,  has been discovered  experimentally \cite{aliceHBT}.

Now the hydrodynamic-based approach to ultrarelativistic nucleus-nucleus collisions becomes the ``standard model'' for these processes. Note, however, that the details of this approach did undergo essential modifications during the last decade and even now they are different in different models. The first attempts to  describe successfully $v_2$ coefficients \cite{Heinz-Kolb} at RHIC were based on the perfect hydrodynamics with an equation of state (EoS) corresponding to a first order phase transition. However, a simultaneous description of $v_2$, particle spectra and femtoscopy scales within the unified initial conditions collided with an invincible problem. Only in recent years have the main factors that make it possible to describe simultaneously the spectra and femtoscopic scales become clear. They are \cite{Sin2006, hkm1, sinyukov-iflow, Bron-flow, Pratt_0}: a relatively hard EoS because of a crossover transition (instead of the 1st order one) between quark-gluon and hadron phases, the presence of the prethermal anisotropic transverse flow developed to thermalization time, an ``additional portion'' of the transverse flow caused by the shear viscosity effect and
fluctuations of the initial conditions. An account of these factors gives the possibility to describe well the pion and kaon spectra together with the femtoscopy RHIC data within a realistic freeze-out picture with a gradual decay of the fluid into observed particles \cite{hkm2}.

In fact, at the moment there is no real unique model of heavy ion collisions. Different hydrodynamic-based approaches use different initial conditions  as well as different final state treatments for the matter evolution, different EoS, etc. In this paper we make the  next step towards converging to the ``standard model''.  We analyze a possibility for a uniform description of the soft observables in the hydrokinetic model of A+A collisions at RHIC and LHC, namely, a simultaneous description of the hadronic yield; pion, kaon, and proton transverse momentum spectra at different centralities; $v_2(p_T)$ coefficients; and femtoscopy scales in A+A collisions at RHIC and LHC. For this aim we use the hydrokinetic model (HKM) \cite{hkm, hkm1, hkm2} in its hybrid version (hHKM) \cite{QM2011}, which allows one to apply hydrodynamics also at the late  non-equilibrated stage of gradual system decay. It gives the possibility to switch correctly to the ultrarelativistic quantum molecular dynamics (UrQMD) cascade at any spacelike hypersurface, in particular, at an isochronic one. We can compare the results also with a pure hybrid model where the hydrodynamics and hadronic cascade are matching just at a non-space-like hypersurface of the chemical freeze-out or on the isochronic hypersurface. The goal of this paper is to fix the HKM by applying it to describe the known experimental data at RHIC and LHC and based on this to predict still unmeasured observables. The results presented here are correlated with similar studies performed in other hydro based models, different from hHKM in various aspects \cite{EPOS2, bozekVisc, BozekLHC}.

\section{Model description}
\subsection{Initial conditions}
All the results presented in this work are calculated for the midrapidity region with the approximation of longitudinal boost-invariance. The  hydrokinetic code \cite{hkm2} is developed now to simulate the full 3+1D matter evolution  and it allows one to analyze the central, noncentral, and peripheral heavy ion collisions.
The formation of the flow anisotropy depends on the initial eccentricity $\epsilon$  which is defined by the centrality of collisions. Even at the maximal centrality the event-by-event fluctuations of the initial conditions (IC) lead to some nonzero effective eccentricity. These fluctuations appear due to chaotic distribution of nucleons (or local color charges) in colliding nucleus and, as the result, the nucleon scattering positions fluctuate from event to event. In the hydrodynamic picture such initial eccentricity of the (mean) energy density profile transforms into a flow anisotropy even at zero impact parameter. The effect persists for non-central heavy ion collisions, where it leads to some systematic increase of mean initial eccentricity.

In this work, we employ the Monte-Carlo Glauber (MC-Glauber) ICs.

\subsubsection{The Glauber initial transverse profile}
The mean transverse density of nucleons in nuclei is defined as
\begin{equation}\label{T}
 T({\textbf x_T})=\int\limits_{-\infty}^{\infty}\frac{a}{\exp[(\sqrt{x_L^2+x_T^2}-R_A)/\delta]+1} dx_L,  
\end{equation}
where $A$ is the atomic number; for Au+Au collisions one uses $A=197$, then $R_A=1.12A^{1/3}-0.86A^{-1/3}\approx6.37 $fm, $\delta=0.54$ fm. Constant $a$ is defined by the normalization condition. In the so-called Glauber model the multiplicity (or entropy) produced in collisions of two nucleus  is defined, roughly, by the number of participating nucleons contained in the overlapping distributions (\ref{T}) for each nuclei shifted by the value of the impact parameter. In the MC-Glauber approach,  instead of utilization of the mean participant density (\ref{T}), one starts the Monte-Carlo procedure with Eq. (\ref{T}) as the distribution function of the random variable-nucleon number at point $x_T$. The finite size of the nucleon can also be taken into account; this slightly modifies the average nucleon distribution compared to Eq. (\ref{T}). The collision criterion for a pair of nucleons in each event is based on the value of nucleon-nucleon cross section $\sigma_{NN}$ at the corresponding collision energy. The nucleon-nucleon collisions result in deposition of a certain amount of multiplicity (entropy) to different cells in a transverse plane. The contributions to multiplicity (entropy) from the ``hard'' elementary collisions and from the ``soft'' ones are different: The former are proportional to the number of binary collisions  while the later are associated with the number of wounded nucleons, or participant number \cite{Kharz_0}. The GLISSANDO code \cite{gliss} allows one to calculate the $x_T$ distribution of both numbers utilizing the MC-Glauber procedure. We use this generator and suppose that  the initial entropy profile in the transverse plane at midrapidity is proportional to a linear combination of the density of wounded nucleons and that of binary collisions:
\begin{equation}
 s(x_T)=C(\frac{1-\delta}{2}\frac{dN_\text{w}}{d^2 x_T}+\delta\frac{dN_\text{bin}}{d^2 x_T})
 \label{Glauber}
\end{equation}
where $C$ is the normalization constant, $\delta=0.14$ for top RHIC energy and $\delta=0.08$ for 2.76 TeV LHC energy are fixed from the description of centrality dependence of multiplicity of all charged hadrons at midrapidity $dN_\text{ch}/d\eta$ in a hydro+cascade model \cite{delta,delta2}.

The MC-Glauber model requires the (inelastic) nucleon-nucleon cross section $\sigma$ and the number of the nucleons $N_\text{a,b}$ in the colliding nuclei as input. We fix this to ($\sigma=42$ mbarn, $N_\text{a,b}=197$) for Au-Au collisions for top RHIC energy, and ($\sigma=64$ mb, $N_\text{a,b}=208$) for the LHC case.
Different centrality classes are defined via the corresponding cuts on the number of participants, as shown in Table \ref{table-cuts}.
The other choice is to introduce cuts on impact parameter instead (which is associated with a cut on average number of participants); however, we found that it does not lead to visually different profiles.

\begin{table}
\begin{tabular}{|l|l|l|l|l|l|l|}
\hline
 c [\%] & 0-5 & 5-10 & 10-20 & 20-30 & 30-40 & \dots \\ \hline
 AuAu, RHIC & $>$328 & 328-279 & 279-201 & 201-143 & 143-98 & \dots \\ \hline
 PbPb, LHC & $>$360 & 360-309 & 309-225 & 225-161 & 161-111 & \dots
\\
\hline
 \end{tabular}
\caption{Cuts on the number of participating nucleons from MC-Glauber model, corresponding to different centrality classes for Au-Au and Pb-Pb collisions at top RHIC and 2.76 TeV LHC energies, respectively.}
\label{table-cuts}
\end{table}

\begin{figure}
\includegraphics[width=0.23\textwidth]{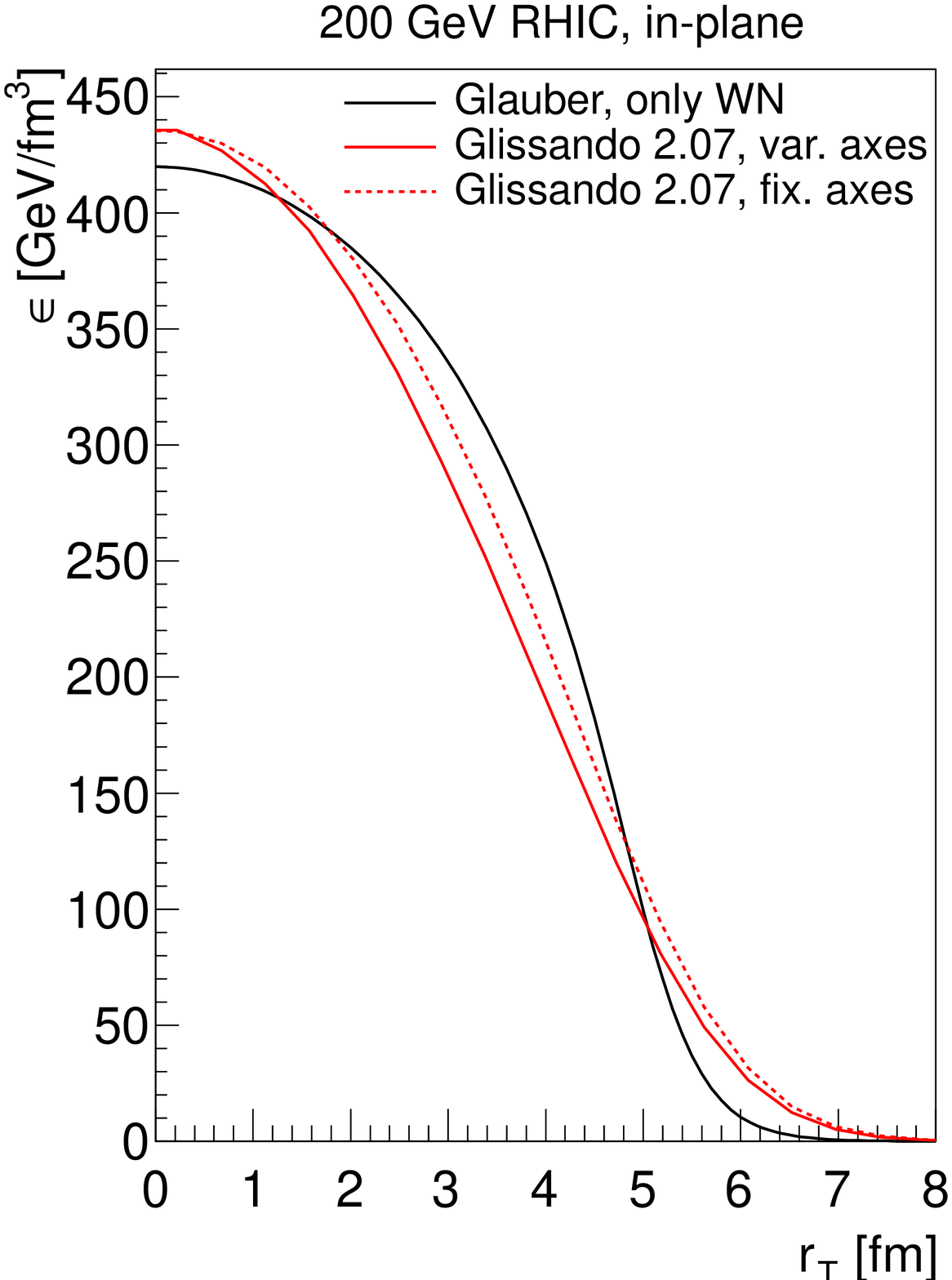}
\includegraphics[width=0.23\textwidth]{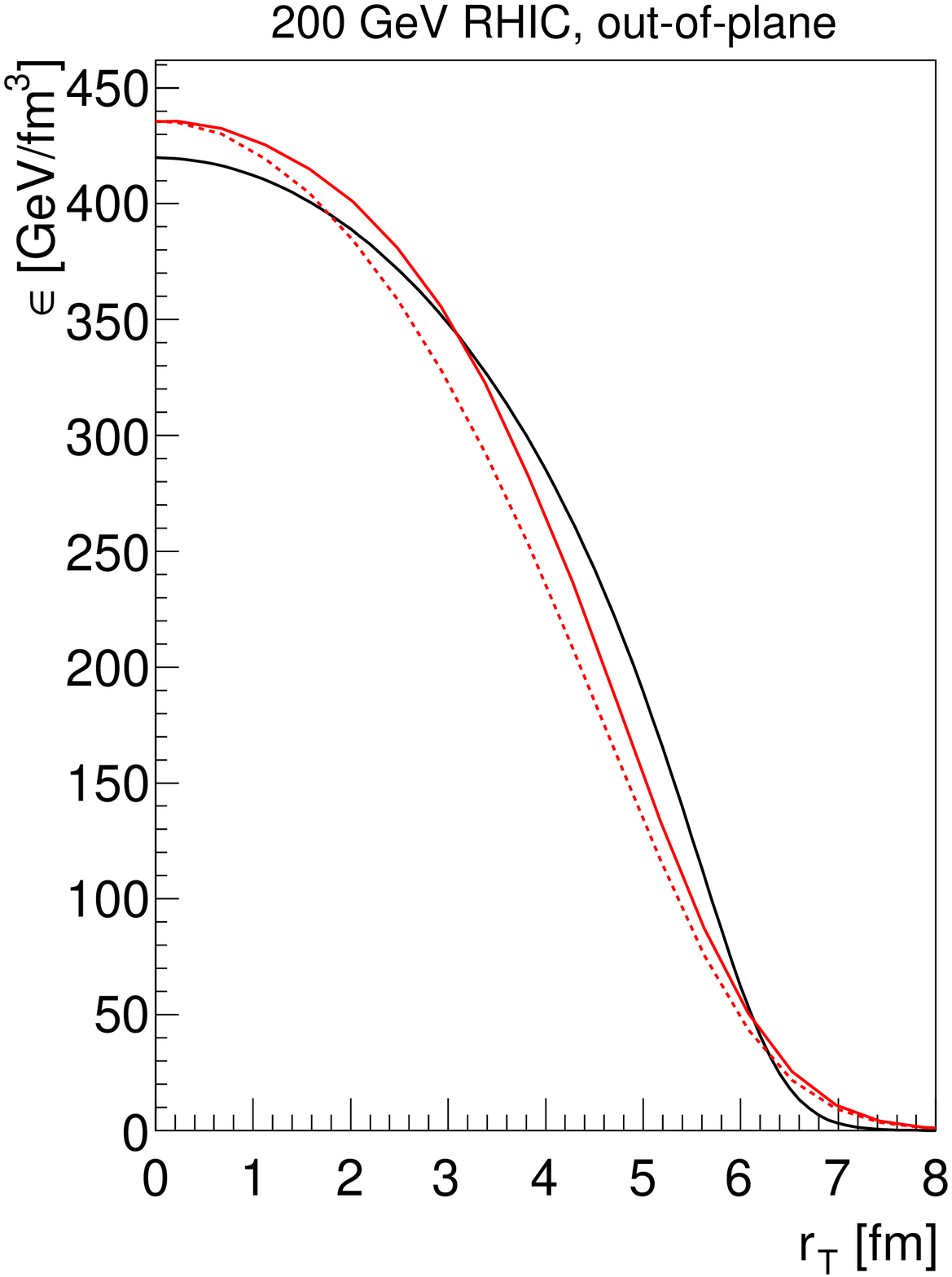}
\caption{(Color online) The initial energy density profile along transverse $x$/in-plane (left) and $y$/out-of-plane (right) directions in the classical Glauber model (black curve) versus the MC-Glauber model with variable-axes analysis (solid red curve) and fixed-axes analysis (dashed red curve). For the MC-Glauber case, $\delta=0.14$ is used in Eq. (\ref{Glauber}).}\label{fig-ic}
\end{figure} 

Because we do not employ event-by-event hydrodynamics here, we average the profiles from a large-enough ensemble of events for each centrality class. The statistical fluctuations tilt in each event the principal axes of the ellipse of inertia and shift the center of mass relative to the reaction-plane coordinate system. To account for this effect, we superpose the principal axes by rotation and re-centering of each initial distribution and after that take averages over the ensemble of events (so-called variable geometry analysis, also implemented as an option in GLISSANDO code).
This gives us nonzero initial eccentricity as a consequence of fluctuations, even for simulations with zero impact parameter. The resulting averaged distribution is then associated with the initial entropy density for hydrodynamic evolution.

In Fig. \ref{fig-ic} we show the resulting energy density profiles for 0-5\% centrality collisions at the top RHIC energy from the MC-Glauber model, compared to the profile from classical Glauber. Note that the second term in Eq. (\ref{Glauber}) corresponding to a contribution from binary collisions leads to a squeezed energy-density profile, which results in the increased radial flow in the system. At the same time, additional initial eccentricity (compared to the classical Glauber model) is gained mostly by performing the variable geometry analysis.

\subsubsection{Initial time and transverse flow}
In the hydrodynamic approach the initial entropy (or energy) density profile should be attributed to the initial  time $\tau_i$ when (partial) thermalization is established. At this moment the system gains  anisotropic prethermal transverse flow that is caused by the transverse finiteness of the system \cite{Sin2006,sinyukov-iflow}. Typically the flow is almost nonrelativistic, $v_T\approx y_T$, and for the Gaussian initial profile for freely expanding fields or partons, which are suddenly thermalized at the time $\tau_i$, is proportional to the transverse coordinate $r$ and inversely proportional to the homogeneity lengths squared  \cite{sinyukov-iflow}. If the homogeneity length is directed along an axis that is tilted by angular $\phi$  to the in-plane axis $x$,
then one can write it in the form
\begin{equation}\label{iflow}
 y_T=\alpha\frac{r_T}{R^2(\phi)}
\end{equation}

Naturally, the developing of the prethermal flows are different from that in hydrodynamic or free-streaming scenarios  because of the different mechanisms of the flow generation at the prethermal and thermal stages. The initial velocity profile is also different from Eq. (\ref{iflow}) because of non-Gaussian initial energy distributions. The utilization of the method  \cite{AkkSin} to the thermalization processes in A+A collisions at RHIC and LHC, which are neither hydrodynamical nor suddenly thermalized free-streaming ones, is in progress now. However in this work we use a very rough  approximation of ideal hydrodynamic evolution  for the prethermal stage. This stage should start from the very initial time of the collision process, just after the c.m.s. energy in overlapping region of colliding nucleus exceeds their binding energy, say at $\tau_{0}=0.1$ fm/$c$;  this time has to be used in a consistent approach.

At initial time $\tau_{0}=0.1$ fm/$c$ there is no collective transverse flow and our basic analysis is related just to this case. In addition, in Sec. \ref{sect:iflow} we demonstrate  the possible corrections caused by the modifications of the final radial and elliptic hydrodynamic flows, which are qualitatively similar to a corresponding effect caused by the shear viscosity.

\subsection{Hydrodynamic evolution and equation of state in the chemically equilibrated zone}

The quark-gluon plasma and hadronic gas are supposed to be in complete local equilibrium above the chemical freeze-out temperature $T_{ch}$ with EoS at high temperatures as in the lattice QCD. With the given ICs, the evolution of thermally and chemically equilibrated matter is described with the help of the ideal hydrodynamic approximation. The latter is based on 3+1D (in the general case) numerical hydrodynamic code, described in Ref. \cite{hkm2}. For this stage of evolution we use the latticeQCD-inspired EoS of the quark-gluon phase \cite{laine} together with corrections for small, but nonzero baryon chemical potentials \cite{karschMu,hkm2}, matched with chemically equilibrated hadron-resonance gas via crossover-type transition.
The hadron-resonance gas consists of all ($N=329$) well-established hadron states made of $u$,$d$,$s$ quarks, including $\sigma$ meson [$f_0$(600]. Quantitatively, the EoS table used is not visually different from widely used \texttt{s95p} EoS parameterization \cite{huovinenEoS,songEoS}, which is based on results from 
the hotQCD collaboration \cite{chengEoS,bazavov}.

\subsection{Hydrokinetic and UrQMD cascade stages}

At the temperatures below $T_{ch}$ the system loses chemical and thermal equilibrium and gradually decays. In the hydrokinetic approach  the dynamical decoupling is described by the particle escape probabilities in inhomogeneous
hydrodynamically expanding systems in a way consistent with the kinetic equations in the relaxation-time approximation
for emission functions \cite{hkm1,hkm2}. The method allows one, in principle, to take into account the back reaction of the particle emission on the system evolution, which is, however, a fairly complicated technical procedure that is equivalent to viscous hydrodynamics only at small deviations from local equilibrium, and we do not apply it. However, without accounting for the back reaction, a serious corrections of the emission function at large times are needed. It can be shown that the hydrokinetic approach without such corrections results in overestimated effective temperature of proton spectra at RHIC energy, and not enough rise of interferometry radii and volume from top RHIC to LHC energies \cite{QM2011}.
To solve this problem we do not employ in this work the hydro-kinetic approach at large times until free-streaming of finally produced particles. Instead of this we switch over a hydrokinetic evolution of continuous medium to an evolution of particles within UrQMD hadronic cascade \cite{urqmd}. The model which matches the HKM and UrQMD we call hybrid HKM (hHKM). 

\subsubsection{The Cooper-Frye switching over hydrodynamics to UrQMD cascade }
In the present analysis, we compare the two different approaches to match  hadrodynamic and hadronic cascade stages. The first one is standard and corresponds to sudden transition from the hydrodynamic regime to the UrQMD cascade.  In this approach, the distribution of $i$-th sort of hadrons at switching hypersurface is expressed through hydrodynamic and thermodynamic variables $u(x), T(x), \mu_i(x)$ by the Cooper-Frye formula \cite{cooperFrye}:
\begin{align}\label{cooperFryeEq}
 p^0 \frac{d^3N_i}{p_T dp_T d\phi_p dy}=\int\limits_{\sigma_\text{sw}} &f_i^\text{l.eq.}(p\cdot u(x), T(x), \mu_i(x))\cdot \nonumber\\
 & \cdot\theta(p^\mu n_\mu) p^\mu d\sigma_\mu
\end{align}
where $n^\mu$ is the hypersurface normal unit four-vector with a positively defined zero component $n^0>0$. The obtained distributions after the Monte Carlo procedure serve as an input for UrQMD cascade. It has been known for a long time that such a prescription leads to inconsistencies \cite{sinyukovCorrCFP, Bug_0}, if the switching hypersurface contains
the non-space-like sectors, and so should be modified to exclude
formally negative contributions to the particle number of the particles that move towards the fluid. A cut of the negative contributions \footnote{This cut is implemented by the additional factor $\theta(p^\mu n_\mu)$ in the right-hand side of Eq. \ref{cooperFryeEq}, where $n_\mu$ is the normal vector to the switching hypersurface \cite{bugaev}, or by more sophisticated factors preserving particle number flow through the hypersurface \cite{sinyukovCorrCFP}} leads in its turn, to loss of causality and distorts  hydrodynamic evolution. In addition, the particles interacting within UrQMD can interact also with fluid, and this opacity effect has to be taken into account, but usually not.
So, despite its technical simplicity, the Cooper-Frye prescription should be applied with caution. 

The method of direct matching of  hydrodynamics and UrQMD cascade  with Cooper-Frye prescription at the chemical freeze-out hypersurface we call traditionally a ``hybrid''  model.  The chemical freeze-out isotherms typically contain  non-space-like parts that  can affect the results for some observables, and we shall check how large the effect is, comparing the results with ones obtained at the hydrokinetic matching at an isochronic hyperesurface. 
  
If the matching hypersurface that separates (pure) hydrodynamics and UrQMD cascade is isochronic, we call this model a ``hybrid-isochronic'' one. In such a model the above-mentioned problems with non-space-like hypersurface are absent. On the other hand,  if one uses an isochronic hypersurface to match the cascade and hydrodynamic stages, it can contradict to the obvious expectation that at the periphery of such a hypersurface any local-equilibrium input to the cascade stage is  not correct. The peripheral regions are spatially and temporally rather far from the 
freeze-out isotherm, have rather small temperatures and thus cannot be described hydrodynamically. We compare the results of the ``hybrid-isochronic'' model with the hHKM approach because the latter allows one to match UrQMD with a decaying hydrodynamic system, having nonequilibrated particle distributions, at isochronic hypersurfaces and in this way to overcome all mentioned problems.    

\subsubsection{Matching of the hydrokinetic stage and hadronic cascade}

To construct the matching between hydrokinetics and UrQMD  one calculates the distribution function for each sort of hadron at the switching hypersurface. Using the technique of the integral form of the Boltzmann equation \cite{hkm1, hkm2}, one can express the UrQMD input as the collection of all particles that propagate freely  from the points of their last collision or just from the point of their creation and do reach this hypersurface. Because the evolution time is the Bjorken proper time $\tau$, the distribution is expressed naturally in hyperbolic coordinates. Then the input for UrQMD is constructed as:    
\begin{widetext}
\begin{align}\label{f}
	&f_i(\tau,\vec x,\vec p)=f_i^{l.eq.}(\tau_0,\vec x^{(\tau_0)}(\tau),\vec p)
	\exp{\left(-\int\limits_{\tau_0}^\tau \tilde R_i(s,\vec x^{(s)}(\tau),\vec p)ds\right)}+ \\
	&\int\limits_{\tau_0}^\tau d\lambda\left[
	f_i^{l.eq.}(\lambda,\vec x^{(\lambda)}(\tau),\vec p)
	\tilde R_i(\lambda,\vec x^{(\lambda)}(\tau),\vec p)+\tilde G_i^{decay}(\lambda,\vec x^{(\lambda)}(\tau),\vec p) \right]
	\exp\left(-\int\limits_{\lambda}^\tau [\tilde R_i(s,\vec x^{(s)}(\tau),\vec p)+\tilde D_i(s,\vec x^{(s)}(\tau),\vec p)]ds\right) \nonumber
\end{align}
\end{widetext}
here $\vec x=\{\vec r_T, \theta\}$, $\tau=\sqrt{t^2-x_L^2}$ is the proper time, $m_T=\sqrt{m^2+p_T^2}$
is the transverse mass, $\theta=\eta-y$, $\eta$ is the space-time rapidity,
and $y$ is the particle rapidity; notation
\begin{align}
&x^{(s)}(\tau)=\{\theta^{(s)}(\tau),\vec r_T^{(s)}(\tau)\} \\
=&\{ sh^{-1}\left(\frac \tau s sh(\theta)\right), \vec r_T-\frac{p_T}{m_T}(\tau ch\theta-\sqrt{s^2+\tau^2 sh^2\theta}) \}
\end{align}
is used.

The different terms in Eq. (\ref{f}) are as follows: $\tilde R_i(\lambda,\vec x,\vec p)=\frac{p_\mu u^\mu}{m_T \cosh\theta} R^*_i(p,T)$ is the collision rate of the $i$-th sort of hadrons with the rest of particles, $\tilde G_i^{decay}(\lambda,\vec x,\vec p)$ is an income of particles into the phase-space point owing to resonance decays, and $\tilde D_i(\lambda,\vec x,\vec p)=\frac{m}{p^0 \cosh\theta}\Gamma_i$ is the decay rate of a given resonance. To calculate the collision rates, we assume meson-meson, meson-baryon and baryon-baryon cross sections in a way similar to the UrQMD code \cite{urqmd}.

The hydrokinetic approach allows one to build the (nonequilibrium) distribution function accounting for the above-mentioned effects on any hypersurface $\sigma_\text{sw}$ that is situated behind the one corresponding to the hadronization process. As was discussed above in the consistent approach one should use the space-like hypersurface to switch to the hadronic cascade.  For the present analysis, in the case of hydrokinetic switching we use $\sigma_\text{sw}: \tau=\tau_\text{sw}=const$ hypersurface, which is completely space-like. However, other choices for switching space-like hypersurface are also possible.

THe hypersurface of chemical freezeout, which is the  matching hypersurface in pure ``hybrid'' scenario, is typically non-space-like, so the explicit parametrization of the hypersurface $\tau(\vec r_T)$ can be double-valued, as well as the parametrization $r_T(\tau,\phi)$. To escape multi-valued functions one may utilize these different parametrizations in different sectors of the freeze-out hypersurface. In the case of the pure ``hybrid'' model it leads to the different representations for the distribution functions (\ref{cooperFryeEq}) for MC generation in different sectors of the hypersurface,
\begin{equation}\label{Fgen}
 \frac{d^6N_i}{d\tau d\eta d\phi dy dp_T d\phi_p}=F_i(\tau,\eta,\phi,y,p_T,\phi_p),
\end{equation}
if the explicit dependence $r_T=r_T(\tau)$ is assumed, or
\begin{equation}\label{Fgen-2}
 \frac{d^6N_i}{dr_T d\eta d\phi dy dp_T d\phi_p}=F_i(r_T,\eta,\phi,y,p_T,\phi_p),
\end{equation}
if another form $\tau=\tau(r_T,\phi)$ is used. For the isochronic switching procedure in the hydrokinetic approach, based on Eq. (\ref{f}), only the form Eq. (\ref{Fgen-2}) is used.     

Once the distribution function $F$ on the switching hypersurface is known, we proceed with the standard (for present event generators) method of Monte Carlo event generation for input to UrQMD. First, one needs to calculate the maximum value of the distribution functions and the mean multiplicities of each sort of hadron in one event. Then, in each event we randomly generate the exact number of each sort of particle in a given rapidity interval according to Poisson distribution, such that the average multiplicity over a large set of events yields the mean multiplicity in the hydrokinetic approach at the chosen isochronic hypersurface.
Then coordinates and momenta of each particle are generated randomly according to the distributions (\ref{cooperFryeEq}) or (\ref{f}). To simplify the procedure of generation, the accept-reject algorithm is used, which means the following, e.g. in the case of distribution (\ref{Fgen}): We generate the random set of coordinates $x_r=\{\tau,\eta,\phi\}$ and momenta $p_r=\{y,p_T,\phi_p\}$ and a random variable $a$ in interval $a\in[0\dots F_\text{max}]$, where $F_\text{max}$ is the maximal 
value of distribution (\ref{Fgen}). Then, we compare the value of distribution in the given point $F_r=F_i(x_r,p_r)$. If $a<F_r$, then the particle coordinate and momentum set are accepted; otherwise it is rejected, and the new random iteration is made. In this way,  we generate in each event the coordinates and momenta of all particles with $|\eta_s|<2$ and $|y|<2$ intervals, which are wide enough to study particle production at midrapidity.

The generated set of particles is then used as input for UrQMD code \cite{urqmd}, which computes further elastic and inelastic collisions and decays of unstable hadrons. Finally, the output of UrQMD - which is a set of particle positions and momenta - is analyzed to obtain physical results of the model.

\section{The observables in the model}
The azimuthally averaged transverse momentum spectra,
$$\frac{dN}{2\pi p_Tdp_Tdy}$$
 is calculated in the direct  way from a generated event ensemble. When comparing to experimental data, we retain or neglect the feed-down from weak decays in accordance with experimental procedure.

In the described Monte Carlo procedure of event generation, the elliptic flow coefficients are calculated in the model in a ``standard'' way, according to:
\begin{equation}
 v_2=<\cos 2(\phi_p-\Psi_\text{RP})>=<\frac{p_x^2-p_y^2}{p_x^2+p_y^2}>,
\end{equation}
where the brackets correspond to the mean value over the particles from a given event ensemble (in a given $p_T$ bin, for $p_T$-differential $v_2$). The procedure of reaction plane constructing is described in Sec. IIA1.  

For femtoscopy analysis, we first calculate the three-dimensional two-pion correlation function as a function of relative momentum $\vec q=\vec p_1-\vec p_2$, for each $k_T$ bin, where $k_T=(p_1+p_2)/2$ is the average momentum of the pion pair. Following the experimental cuts (which are somewhat different for STAR, PHENIX and ALICE collaborations), we consider pions in central pseudorapidity region $|\eta|<0.5$. To calculate the correlation function, we use the same technique as in our previous studies in the FASTMC event generator \cite{fastmc}. Namely, the correlation function for bosons in the Monte Carlo procedure is calculated according to
\begin{equation}\label{CFs}
 C(\vec q)=\frac{\sum\limits_{i\neq j}\delta_\Delta(\vec q-p_i+p_j)(1+cos((p_j-p_i)(x_j-x_i))}{\sum\limits_{i\neq j}\delta_\Delta(\vec q-p_i+p_j)}
\end{equation}
where $\delta_{\Delta}(x)=1$ if $|x|<\Delta p/2$ and 0 otherwise, with $\Delta p$ being the bin size in histograms. We decompose the relative momentum $\vec q$ into $(q_\text{out},q_\text{side},q_\text{long})$ projections, and preform analysis in the longitudinal center of mass system (LCMS), where the mean longitudinal momentum of the pair vanishes.

\begin{figure*}[!p]
 \includegraphics[width=\textwidth]{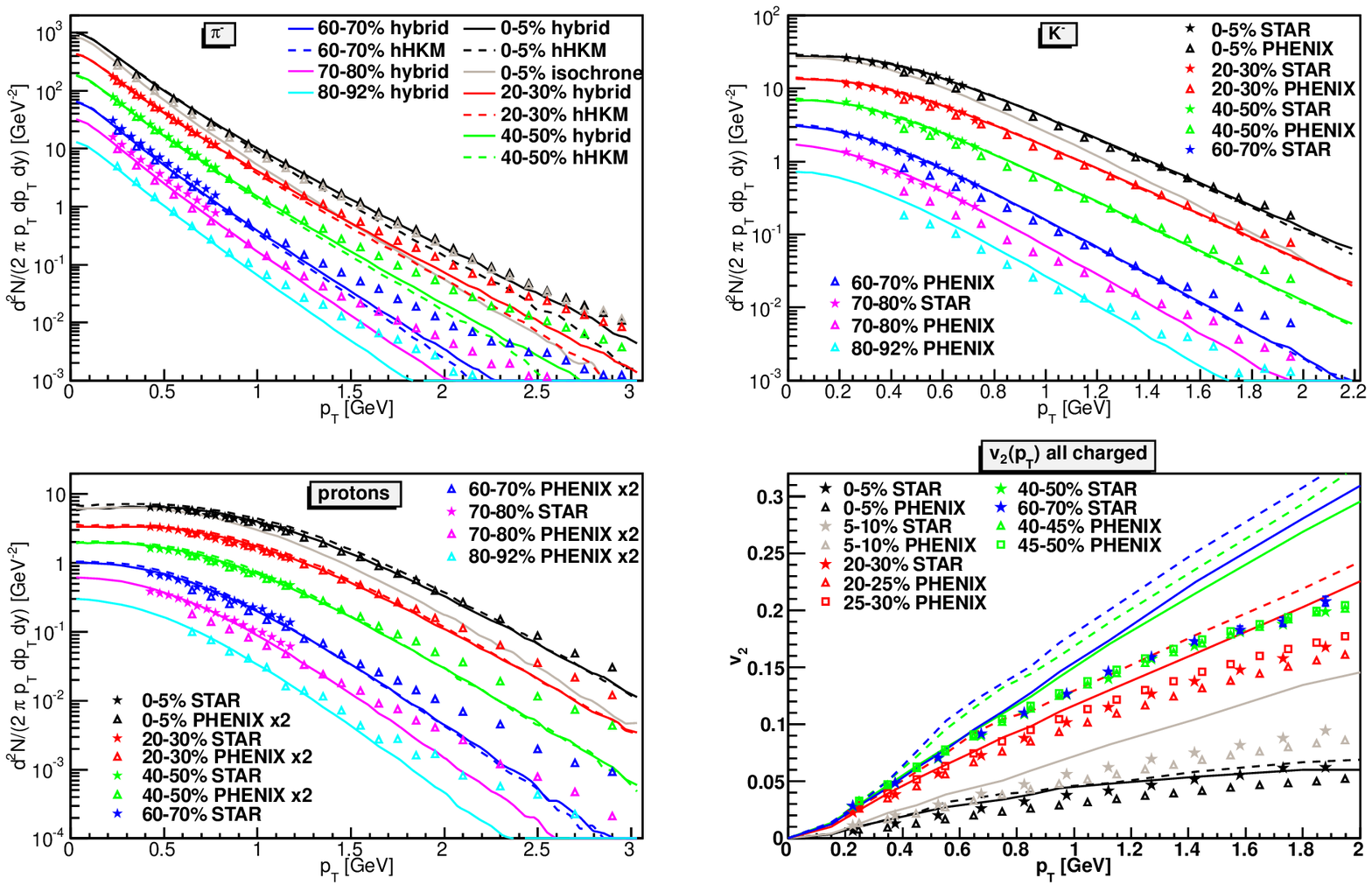}
 \includegraphics[width=\textwidth]{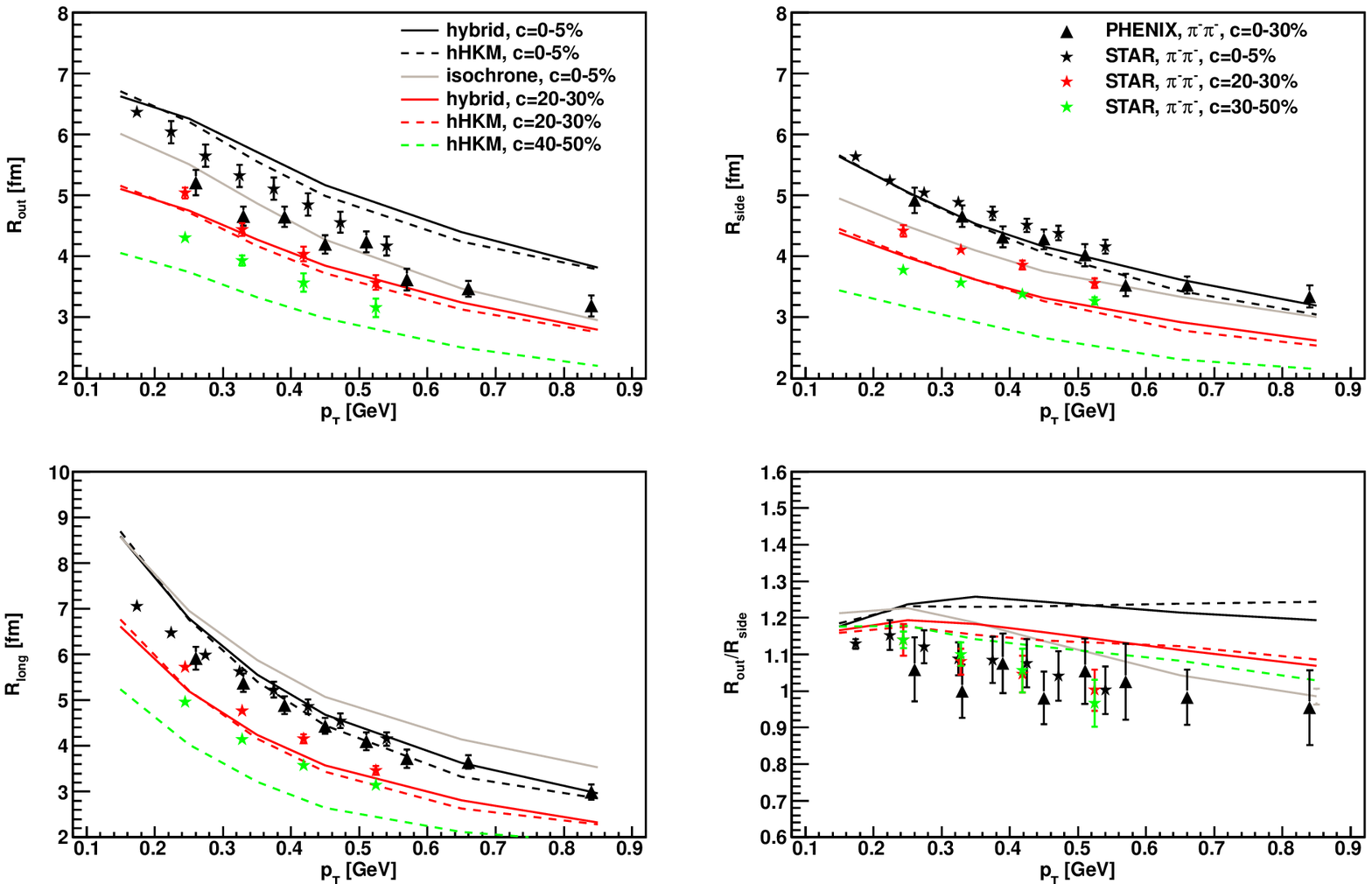}
 \caption{(Color online) $\pi^-$,$K^-$ and proton spectra and elliptic flow of all charged hadrons at midrapidity for 200A GeV RHIC energy and different centrality classes (top) and HBT radii of $\pi^-$ pairs for most central events, calculated in different models and compared to experimental data from the STAR \cite{starSpectra,starv2,starHBT} and PHENIX \cite{phenixSpectra,phenixv2,phenixHBT} collaborations. Note that proton PHENIX spectra is multiplied by the factor 2 (see text). Dashed lines correspond to the hydrokinetic procedure of matching (hHKM), while solid lines stand for the ``hybrid'' model case. The results for the ``hybrid-isochronic'' model are presented for  central $c=$ 0-5\% events with a gray solid line. The ICs used: $\tau_0=0.1$ fm/c with zero initial transverse flow and the MC-Glauber profiles for initial entropy density.}\label{fig-bestfit1}
\end{figure*}

Evaluation according to Eq. (\ref{CFs}) is done with the help of 3D histograms, implemented in ROOT library classes \cite{ROOT}, and two separate histograms are used to calculate the numerator and denominator of (\ref{CFs}), which are divided to get the correlation function. We fit the resulting correlation function with the Bertsch-Pratt parameterization:
$$C(q)=1+\lambda\cdot\exp(-R^2_\text{out}q^2_\text{out}-R^2_\text{side}q^2_\text{side}-R^2_\text{long}q^2_\text{long})$$

Owing to longitudinal boost invariance and approximate azimuthal symmetry for the most central collisions which we consider for the present HBT studies, the cross-terms $R^2_{os}, R^2_{sl}$ and $R^2_{ol}$ are neglected.

\subsection{The results for RHIC}

The model parameters for MC-Glauber ICs are defined as follows. We consider the three switching procedures between hydrodynamics and the hadronic cascade. In the  ``hybrid'' model the chemical freeze-out isotherm with temperature $T_{ch}$ (defined below) serves as the switching hypersurface. In ``hybrid  --  isochronic''  model the switch of hydrodynamics to the UrQMD cascade is implemented at the isochronic hypersurface $\tau_{sw}=$ const, and $\tau_{sw}$ is defined from the requirement that the isochronic hypersurface  touches the chemical freeze-out isotherm at $r=0$: $T(\tau_{sw}, r=0) = T_{ch}$. In the hydrokinetic procedure the switching hypersurface is the same isochronic one but at time $\tau < \tau_{sw}$ the evolution is described by the hydrokinetic but not not pure hydrodynamics. The MC-Glauber initial conditions are attributed to the starting time $\tau_0=0.1$ fm/c in all the scenarios. The basic results are related to zero initial transverse flow. 

The normalization constant $C$ for the initial entropy density in Eq. (\ref{Glauber}) is defined by the all charged particle multiplicity in most central collisions $c=0-5\%$ and is fixed for the given collision energy. The same concerns the parameter $\delta$ that defines the binary collision contribution according to Eq. (\ref{Glauber}). For RHIC $\delta = $0.14. We also fix the parameters of chemical freezeout: $T_\text{ch}=165$ MeV, $\mu_B=28$ MeV, $\mu_S=7$ MeV, $\mu_E=-1$  MeV, according to the analysis of particle number ratios from STAR in thermal model \cite{thermalPBM, thermalBec}, and include contributions from weak decays to the proton spectrum to make correspondence with STAR procedure. Some of the resulting particle number ratios are shown in Table \ref{table-rhic}.
Note that in our hadron table we also include additional resonance states, e.g., $f_0(600)$, $f_0(980)$, $a_0(980)$ and high mass resonances with $m>2$ GeV, followed by the recent compilation from Particle Data Group \cite{pdg}. $f_0(600)$ is observed as a broad resonance structure, with very little knowledge about its decay channels and branching ratios. However, these resonances contribute to both final pion and (less) proton yields \cite{thermalPBM}; thus, 
modifying particle number ratios, in particular, improves (anti)proton yields at RHIC. 

\subsubsection{The yields, spectra, $v_2$ and femtoscopy}

We focus first on the description of the top RHIC Au+Au data in hHKM model, which is shown in Fig. \ref{fig-bestfit1}.

It is known that even for the most central Au+Au collisions STAR and PHENIX proton multiplicities differ by a factor of $\approx2$, pion multiplicities by more than 15\%, and these discrepancies rise from central to peripheral centrality classes, see Fig. \ref{fig-bestfit1}. Within our model this difference cannot be fully reproduced by switching on/off weak contributions only, e.g. to reach lower pion multiplicity at PHENIX one also has to decrease initial energy density.
We choose ICs in the model to reproduce the STAR multiplicities, but compare the results with the spectra and HBT radii measured by both collaborations. Because transverse momentum spectra of identified hadrons are measured in a wide $p_T$ range only by PHENIX, we multiply the PHENIX proton spectrum by the factor 2 to better compare its shape with the STAR points and the hHKM calculations.  From Fig. \ref{fig-bestfit1} 
it is seen that the kaon multiplicity is over-predicted in hHKM for non-central events  if compared to the PHENIX data, which could be a sign of the incomplete equilibration of strangeness in peripheral collisions \cite{incomplTherm}. However, this effect is not seen  when one compares results with the STAR kaon data.
\begin{table}[h]
\begin{tabular}{|l|l|l|l|l|l|l|}
\hline
~ & $N_{\pi^-}$ & $N_{K^-}$ & $N_p$ & $N_{\bar p}$ \\ \hline
STAR & 327$\pm$33 & 49.5$\pm$7.4 & 34.7$\pm$6.2 & 26.7$\pm$4.0 \\ \hline
hHKM & 330 & 47.3 & 29.9 & 20.5 \\
\hline
 \end{tabular}
\caption{Identified hadron yields measured by STAR collaboration \cite{star_spectra} for most central (0-5\%) AuAu collisions at top RHIC energy, compared to hHKM calculations.}
\label{table-rhic}
\end{table}
% We compare with the results of STAR and PHENIX collaborations for transverse momentum spectra for identified hadrons ($\pi^-$, $K^-$ and protons), elliptic flow coefficients and interferometry radii.

\begin{figure}[!htb]
 \includegraphics[width=0.5\textwidth]{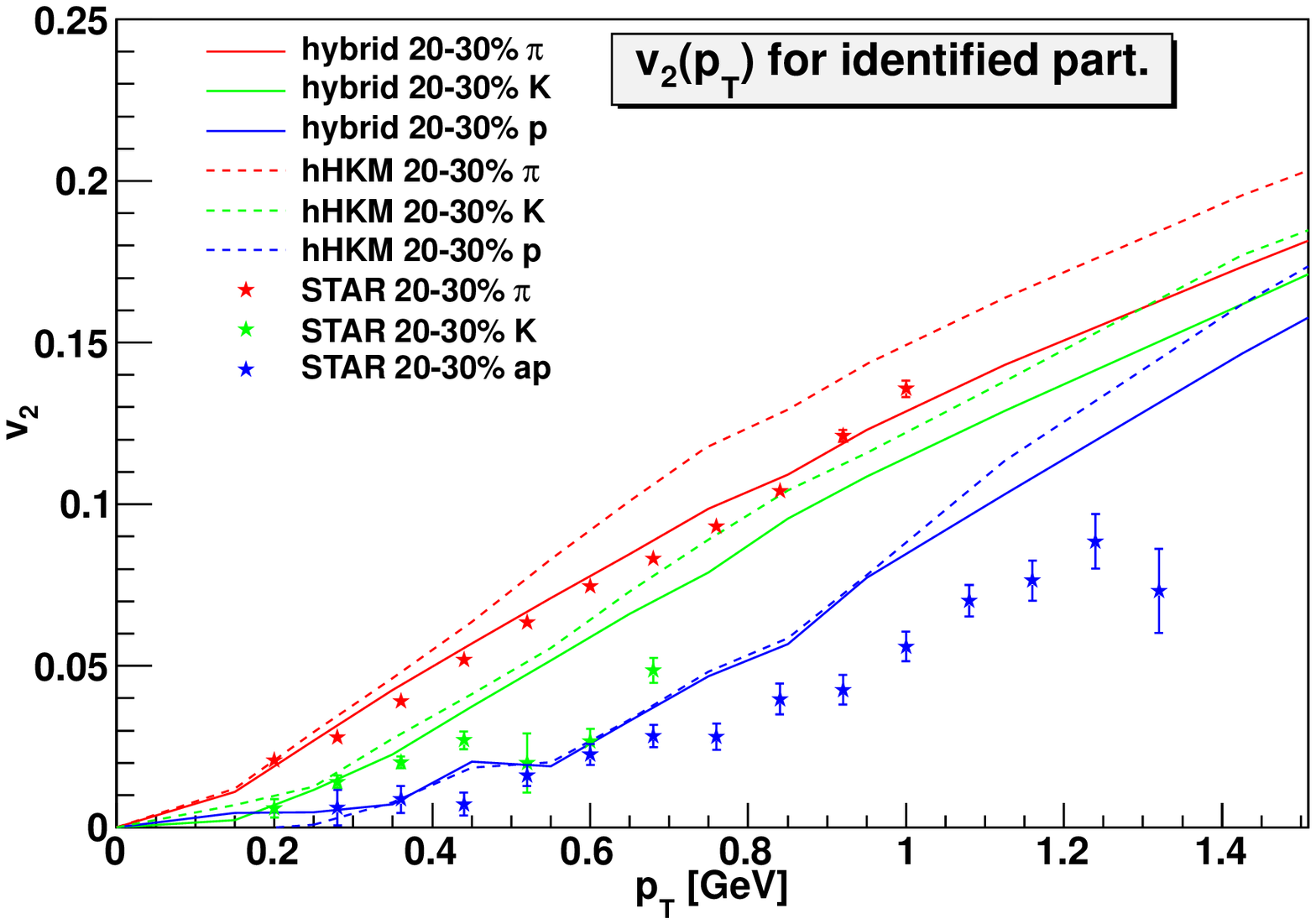}
\caption{(Color online) Elliptic flow of $\pi^\pm$,$K^\pm$ and (anti)protons at top RHIC energy and centrality c=20-30\%, calculated in the hHKM  and ``hybrid'' models with the same parameters as in Fig. \ref{fig-bestfit1} and compared to the STAR data \cite{starv2}}\label{fig-v2ident-rhic}
\end{figure}

In the middle row of Fig. \ref{fig-bestfit1} we compare the elliptic flow coefficients in full hHKM and ``hybrid'' scenarios with the STAR \cite{starv2} and PHENIX \cite{phenixv2} results for all charged hadrons, obtained with event-plane method. In addition, in Fig. \ref{fig-v2ident-rhic} we present elliptic flow of charged pions, kaons and (protons+antiprotons), calculated with the same parameters as for Fig. \ref{fig-bestfit1}, and compared to $v_2\{2\}$ for charged pions, kaons, and anti-protons from STAR (using a two-particle cumulant method).
This calculation is done only for the centrality c=20-30\%, because a considerably bigger set of events must be used for this observable. One can see some overestimate of the data for $v_2$-coefficients, it is because we use ideal hydrodynamics, but not a viscous one.

The comparison of interferometry radii, calculated in the model with the experimental data is shown in the bottom of Fig. \ref{fig-bestfit1}. Note that PHENIX presented its results for the 0-30\% centrality bin, which corresponds to a smaller average multiplicity than the 0-5\% STAR bin, thus PHENIX radii lie slightly below ones calculated by STAR; in our model we observe the same tendency with average ICs for 0-30\% centrality.
  
From Figs. \ref{fig-bestfit1} and \ref{fig-v2ident-rhic} one can conclude	that both the hHKM and the ``hybrid'' models describe the data quite satisfactory, except $v_2$ for very noncentral collisions (where the viscosity should play an important role and reduces elliptic flow) and HBT radii for 40-50\% centrality, which are clearly underestimated. The shear viscosity should also reduce the $R_{out}/R_{side}$ ratio because it enhances the transverse flow. As for the ``hybrid -- isochrone'' model, it fails to describe the shape of pion, kaon and proton transverse spectra, $v_2$, and long-, side- and out- interfereometry radii. The main difference between the first two (hHM, hybrid) and  ``hybrid -- isochrone'' scenarios is that the first two matching procedures do not use the local equilibrium particle distribution functions as input for UrQMD cascade at the space-time regions where the system should be far from equilibrium. The peripheral regions at isochronic hypersurface are spatially and temporally distant from the 
freeze-out isotherm and have rather small temperatures and in this transition area the finite and rapidly expanding system cannot be described hydrodynamically: The free-streaming regime of particle propagation already starts there. However, such a system can be described with the hydrokinetic approach. In what follows we consider only hHKM and ``hybrid'' models.

\begin{figure}
 \includegraphics[width=0.5\textwidth]{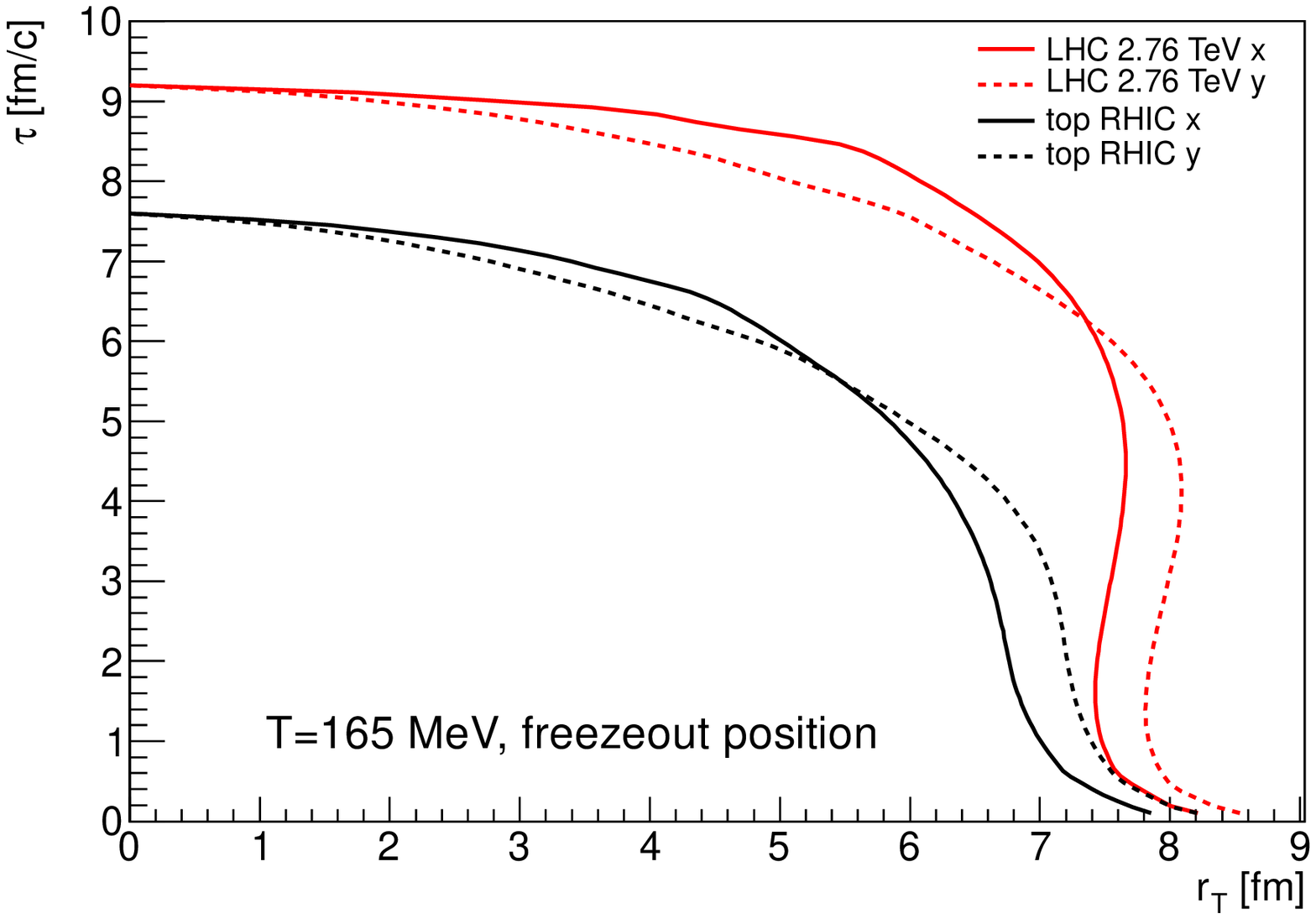}
 \includegraphics[width=0.5\textwidth]{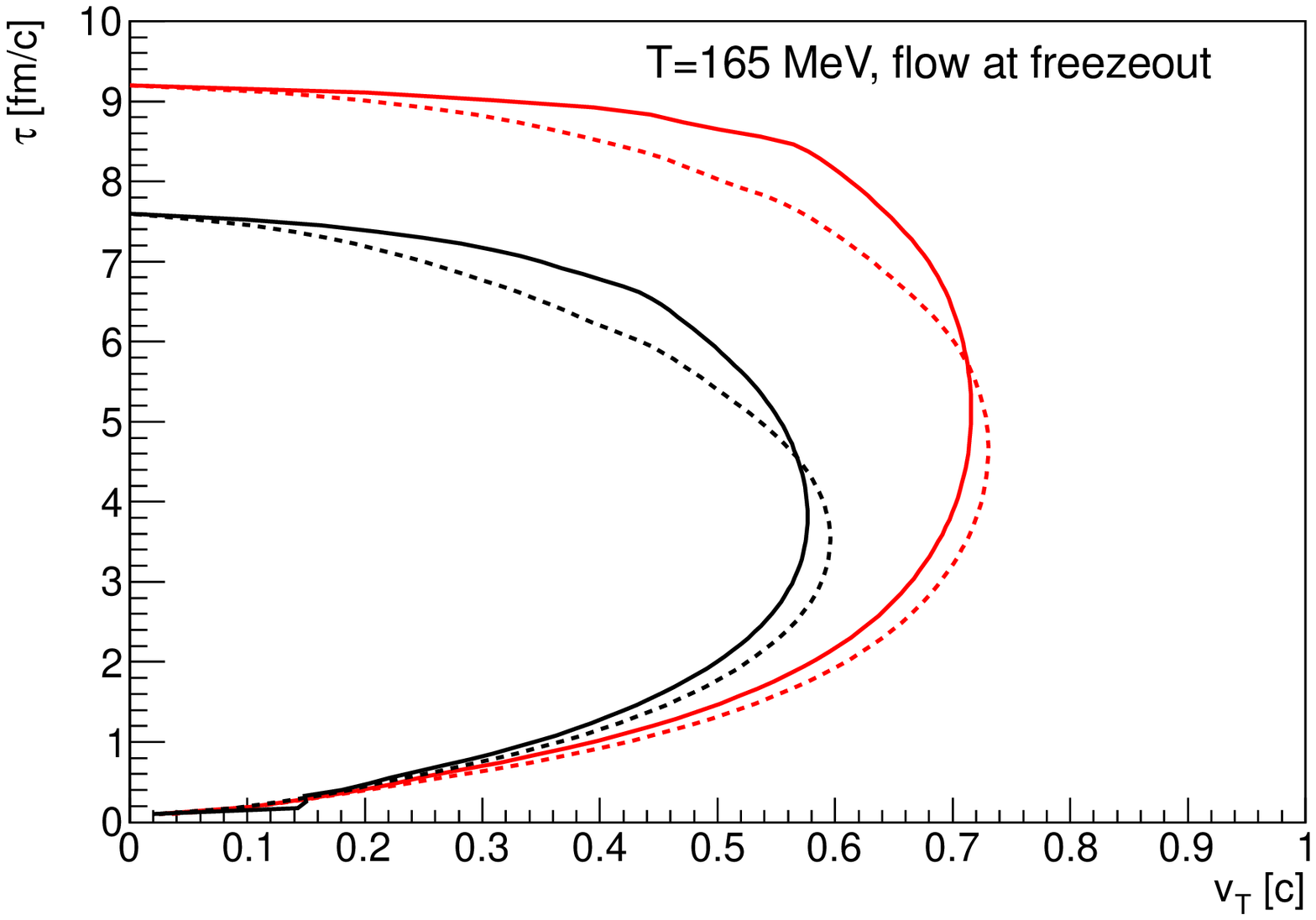}
 \caption{(Color online) Chemical freezeout hypersurface profiles and transverse flows at the chemical freezeout position in the hHKM model. Profiles are made in the in-plane direction (solid lines) and the out-of-plane direction (dashed lines) for top RHIC (black) and 2.76 TeV LHC (red) A+A collisions. Initial conditions are taken from the MC-Glauber model.}\label{fig-lhc-surface}
\end{figure}

\subsection{The results for LHC}

We now move to LHC energy to inspect now well the hHKM and ``hybrid'' models fit to 2.76 TeV LHC Pb+Pb collisions. First we have to account for the corresponding increase of the initial entropy production at the LHC energy by changing the normalization constant $C$ in Eq. (\ref{Glauber}) of the initial energy density profile, the parameter $\delta$ is chosen to be 0.08 \cite{delta2} for LHC energy.  
With bigger initial energy density, hydrodynamic evolution evidently leads to larger effective volume at the chemical freeze-out hypersurface, as well as to bigger transverse flow; see Fig. \ref{fig-lhc-surface}. The second modification is related to the chemical composition at freeze-out: According to approximate particle-antiparticle symmetry at $\sqrt{s}=2.76$ TeV energy, confirmed by ALICE data \cite{alice-spectra}, we set all chemical potentials to zero: $\mu_B=\mu_Q=\mu_s=0$.

\subsubsection{Yields and particle-number ratios}
\begin{figure*}
 \includegraphics[width=\textwidth]{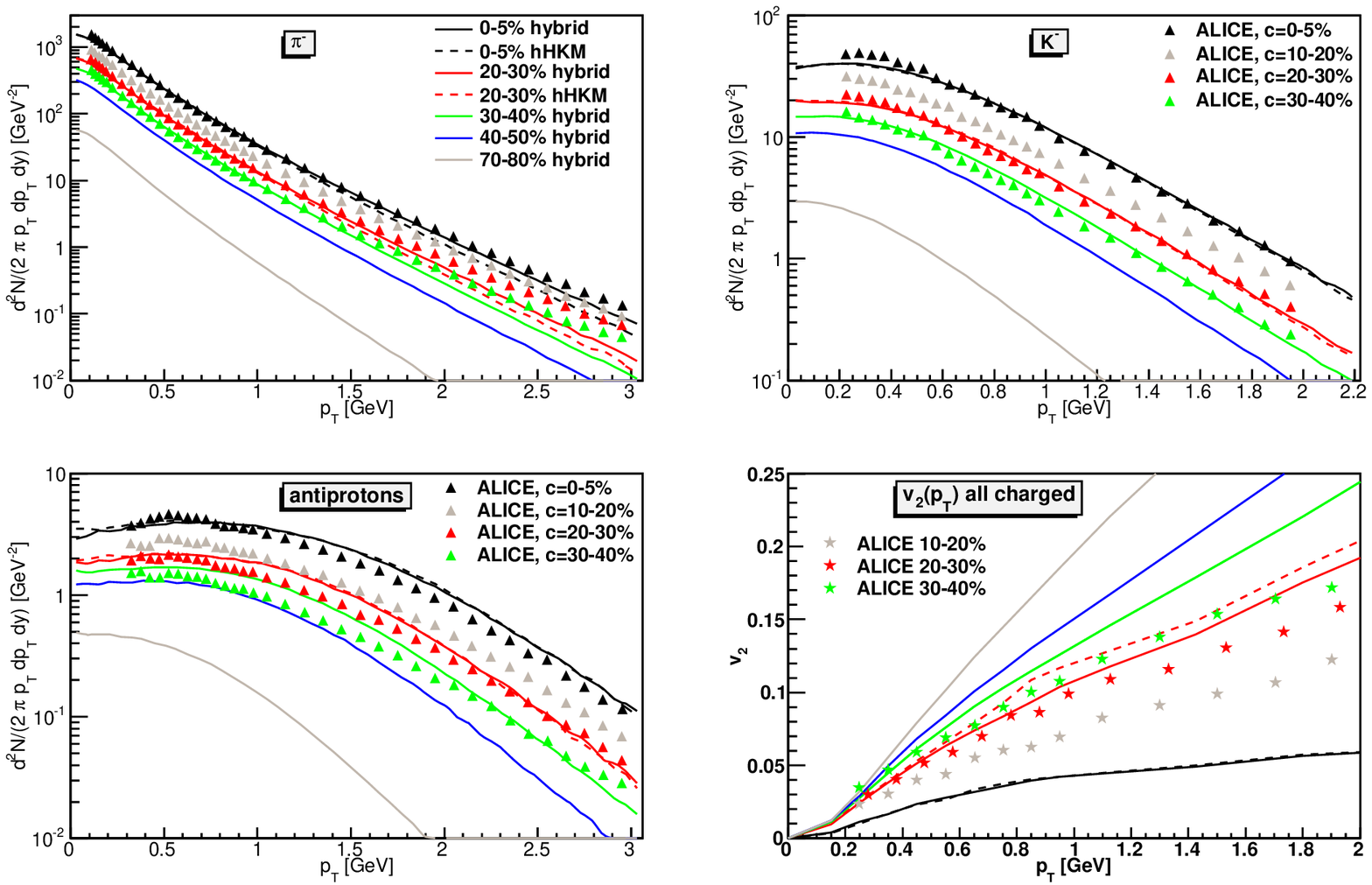}
 \includegraphics[width=\textwidth]{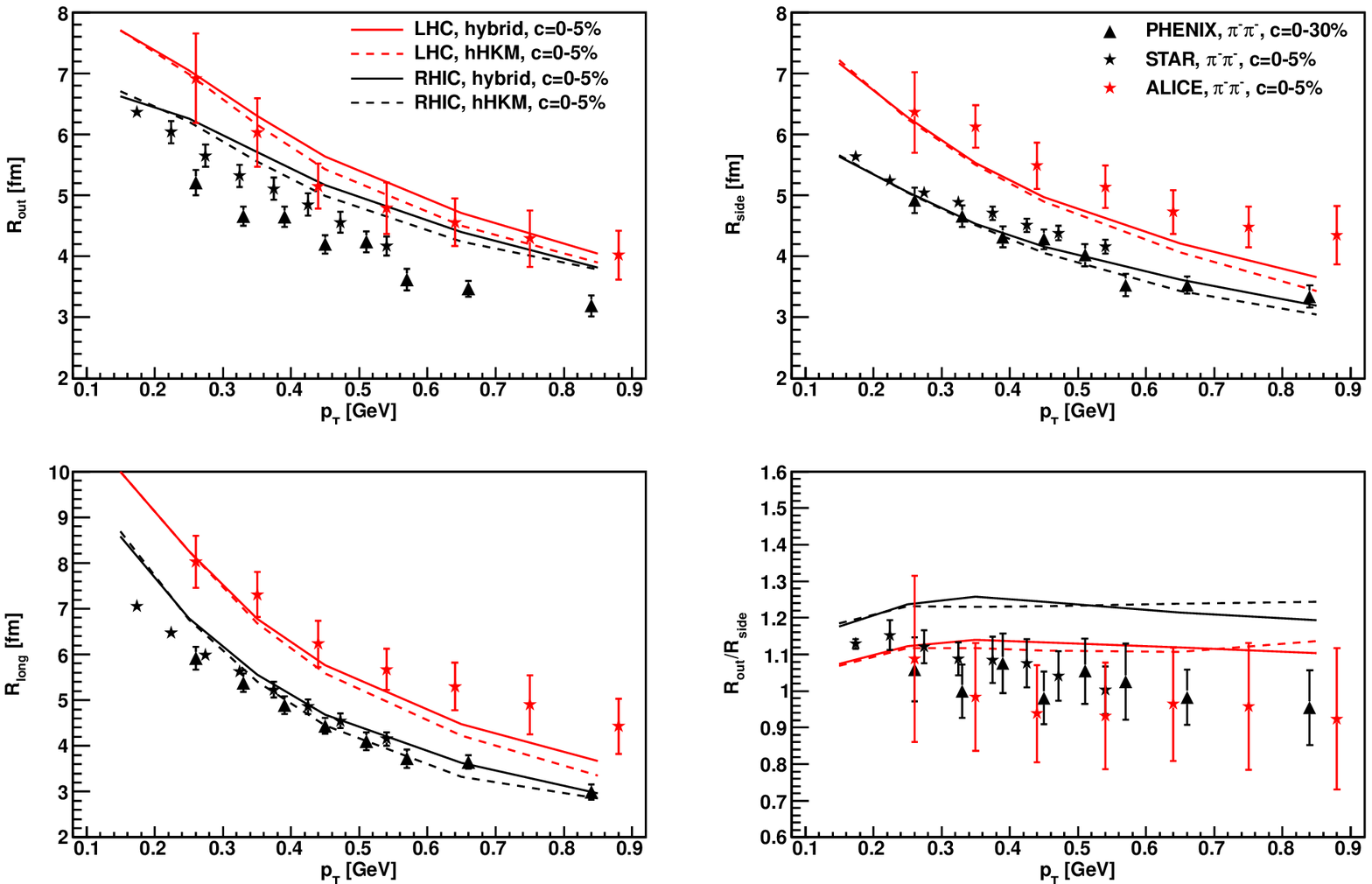}
 \caption{(Color online) $\pi^-$,$K^-$ and proton spectra and elliptic flow of all charged hadrons at mid-rapidity for 2.76 TeV LHC energy and different centrality classes (top) and HBT radii of $\pi^-$ pairs for most central events, calculated in the hHKM model and compared to experimental data from ALICE \cite{alice-spectra,alicev2,aliceHBT}. Dashed lines correspond to hydrokinetic procedure of matching, while solid lines stand for the ``hybrid'' model case. Corresponding HBT radii for top RHIC energy are shown for comparison purposes. Note that calculations for $c=10-20$ \%  are not provided, but the $v_2$-coefficient for $c=0-5$ \% is presented instead as prediction.}\label{fig-lhc}
\end{figure*}

Let us start with the particle-number ratios obtained in the hHKM model for LHC energy. In the simulations we also observe antiparticle/particle symmetry in $\pi^-/\pi^+$, $K^-/K^+$, $\bar p/p$ ratios, which are all close to 1. From the spectra plots in Fig. \ref{fig-lhc} one can conclude that (anti)proton multiplicity at midrapidity in the hHKM model is only slightly overestimated. Indeed, the nontrivial particle number ratios at midrapidity are shown in Table \ref{table-lhc}.
To understand which factors contribute to successful description of particle-number ratios at LHC (and in particular the $p/\pi$ ratio), let us calculate hadron yields in different scenarios of evolution at the post-chemically equilibrated phase, while keeping the same ICs and chemical composition at chemical freeze-out. In this subanalysis we look at the most central collisions, where rescattering effects in cascade -- via UrQMD code -- should be most prominent. From Table \ref{table-lhc} one can see that both pion and proton yields 
are minimal for the ``thermal model'' scenario, where only resonance decays are enabled. Involvement of UrQMD to calculate both elastic and inelastic scatterings (except for baryon-antibaryon annihilation, turned off with \texttt{CTOption(19)=1}) increases somewhat both pion and lambda yields. Finally, switching on $B\bar B$-annihilation suppresses baryon yields, and at the same time increases pion yield, thus lowering the $p/\pi$ ratio to the value $0.052$, which is quite close to the one measured by ALICE \cite{alice-spectra}.
\begin{table}
\begin{tabular}{|l|l|l|l|l|l|l|l|l|}
\hline
~ & $N_\pi$ & $N_K$ & $N_p$ & $N_\Lambda$ & $p/\pi$ & $K/\pi$ & $\Lambda/\pi$ \\ \hline
full & 775 & 123 & 40.5 & 20 & 0.052 & 0.158 & 0.026 \\
full-$B\bar B$ & 716 & 114 & 50.5 & 24 & 0.072 & 0.159 & 0.034 \\
thermal & 679 & 127 & 54 & 20.3 & 0.08 & 0.188 & 0.03 \\
\hline
 \end{tabular}
\caption{Particle multiplicities and particle number ratios, calculated within the hHKM model for most central (0-5\%) PbPb collisions with $\sqrt{s}=2.76$ TeV in differen scenarios of particle production: full scenario (hydro+UrQMD), full-$B\bar B$ (baryon-antibaryon annihilator switched off in UrQMD), and thermal model (kinetic phase with resonance decays only).}
\label{table-lhc}
\end{table}
Thus, one can conclude that annihilation processes in UrQMD are essential for successful reproduction of the $p/\pi$ ratio at the LHC energy. When going to noncentral collisions, the $p/\pi$ ratio slightly increases up to $0.058$ for 20-30\% centrality (consistently with ALICE data), which should be the result of fewer inelastic processes owing to decreased effective volume at the hadronization hypersurface. Because charged hadron density at midrapidity for 20-30\% central PbPb collisions at 2.76 
TeV is close to the one for most central collisions at top RHIC energy, one can also conclude that inelastic processes in the cascade play a similar role also at RHIC, modifying particle number ratios and, in particular, decreasing the $p/\pi$ ratio. It is worth noting that the value of the effect depends on the dynamics of the fireball that defines a duration of the afterburn stage and so can differ at RHIC and LHC energies.

\begin{figure}[!htb]
 \includegraphics[width=0.5\textwidth]{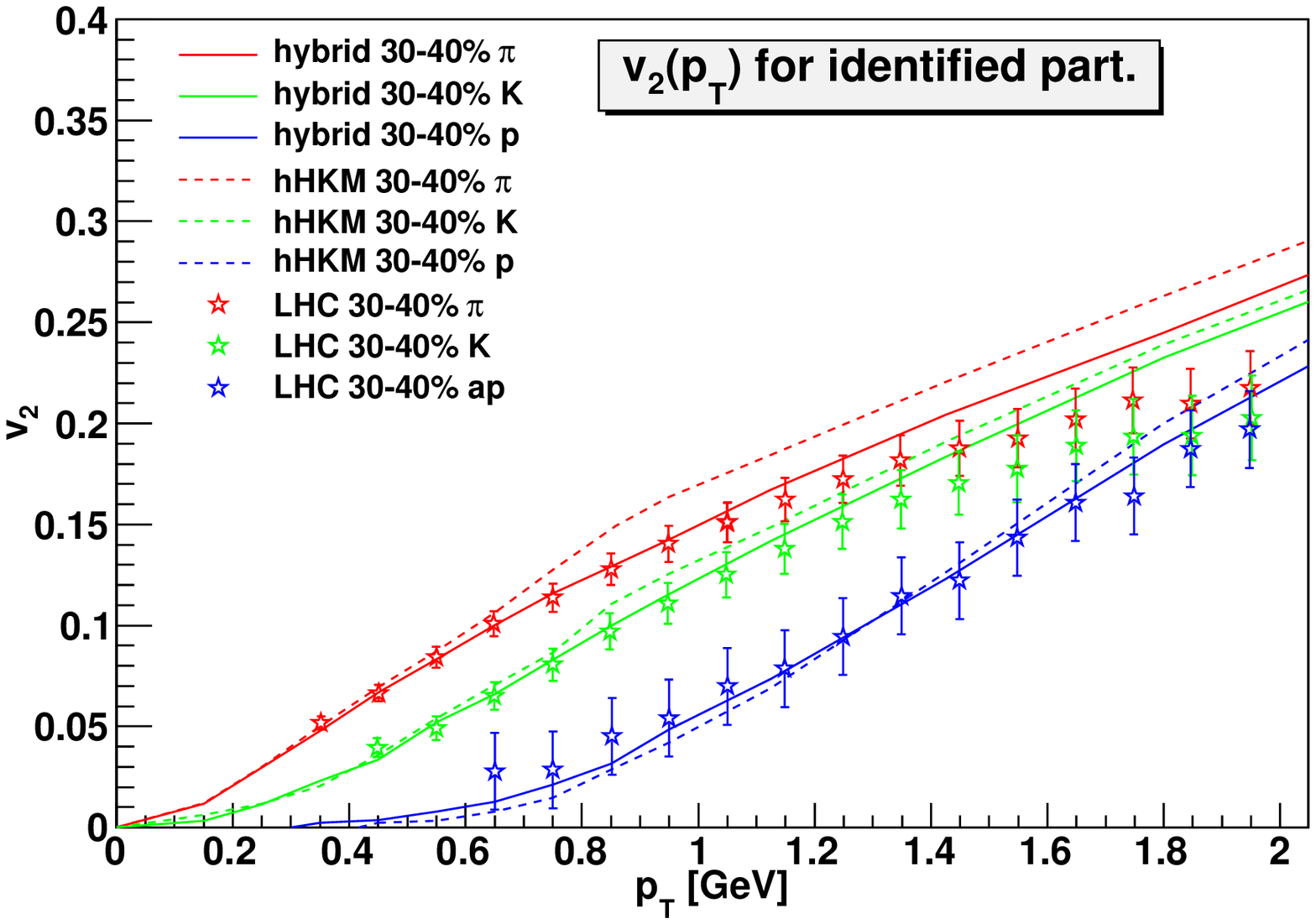}
\caption{(Color online) Elliptic flow of $\pi^\pm$,$K^\pm$ and (anti)protons at top 2.76 TeV LHC energy and centrality c=30-40\%, calculated in the hHKM model with the same parameters as in Fig. \ref{fig-lhc} and compared to ALICE data \cite{alicev2-id}}\label{fig-v2ident-lhc}
\end{figure}

\subsubsection{Spectra, elliptic flow and femtoscopy}
In Fig. \ref{fig-lhc} we show the comparison model/experiment for 2.76 TeV LHC energy. The experimental data are related to elliptic flow coefficients from the four-particle cumulant method, $v_2\{4\}$ \cite{alicev2} and interferometry radii to the most central events measured by the ALICE Collaboration \cite{aliceHBT}. In addition, as for the RHIC case, in Fig. \ref{fig-v2ident-lhc} we compare the elliptic flow $v_2\{2\}$ of charged pions, kaons and antiprotons for the centrality 10-20\%, measured by ALICE \cite{alicev2-id}.
On the last plot, the experimental points are somewhat above the hHKM calculations, and the probable source of the difference is the two-particle cumulant method, which give systematically bigger elliptic flow than the four-particle cumulant method used for the analysis of the elliptic flow of all charged hadrons. To compare with the experimental antiproton spectrum from ALICE, we exclude all feed-downs from weak decays, except for $\bar\Sigma^+$. As a result, we observe some difference 
compared to  ALICE for the mass dependence of the effective temperature (inverse slope) of the transverse momentum spectrum: the resulting antiproton and kaon spectra are too flat in the model, while the  pion spectrum is reproduced much better in a wide $p_T$ region. The reason for such a mismatch could be that the imitation of the viscous effects by the initial transverse flow is not so effective at LHC than at RHIC because of a more protracted (viscous) hydrodynamics stage.
Concerning interferometry results, in hHKM we get systematically lower $R_\text{side}$ and $R_\text{long}$ radii than the ALICE data for larger $p_T$; however, they are within the experimental error bars almost in all $p_T$ regions. Note that the rise of interferometry volume observed by the ALICE collaboration is well reproduced in hHKM (see also Ref. \cite{QM2011}). Because we keep  untouched the main model parameters when passing  from RHIC to LHC energies, except for a general normalization for increased $dN/dy$, contribution from binary collisions, and the baryonic chemical potentials at freeze-out, one can conclude that the soft physics at RHIC and LHC is similar.

\subsection{Possible corrections to the basic  results}\label{sect:iflow}

The above results are presented for perfect fluid dynamics of quark-gluon plasma and nonequilibrium evolution of hadronic gas after the chemical freeze-out. As it is known, quantum theory gives the limitation on the shear viscosity coefficient $\eta$, more exactly, on the minimal ratio $\frac{\eta}{s}=\frac{1}{4\pi}$, where $s$ is entropy density. The inclusion of the shear viscosity in the boost-invariant hydrodynamic models affects the final hydrodynamic flow in the following way: it increases the  radial  and decreases  elliptic flows. 

In this work we do not utilize viscous hydrodynamics. However in this Section we try to demonstrate the influence of the above-mentioned effects  on observed transverse spectra, elliptic flows, and HBT. 
Of course, the conservation laws have to be taken into account to see reliable correlation in modifications of different kinds of observables. For this aim we use our model with modified ICs that affect the final flow in the above-mentioned way. We found that the artificial addition of a small initial transverse velocity with profile  (\ref{iflow}) can serve for this aim. In this case the pressure gradient will drop faster in all directions and this reduces an efficiency of the transformation of the anisotropy of initial pressure gradient into the final flow anisotropy \cite{Sin2006}. Formally, when ${\bf v} \rightarrow {\bf v}+{\bf v}_{rad}$, the anisotropy parameter $\epsilon_v= \left\langle v_x-v_y\right\rangle/\left\langle v_x+v_y\right\rangle$ drops.
\begin{figure*}
 \includegraphics[width=\textwidth]{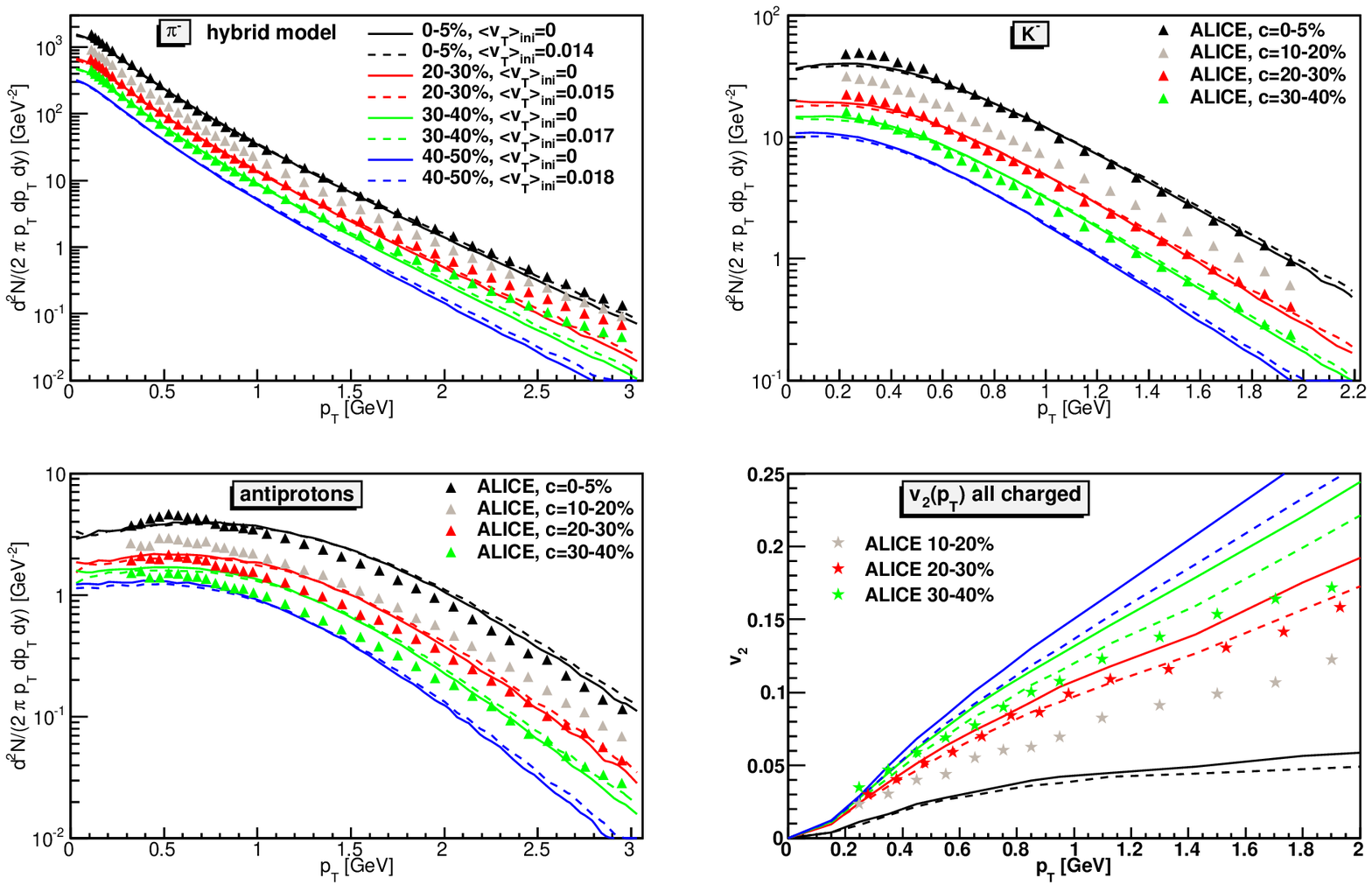}
 \includegraphics[width=\textwidth]{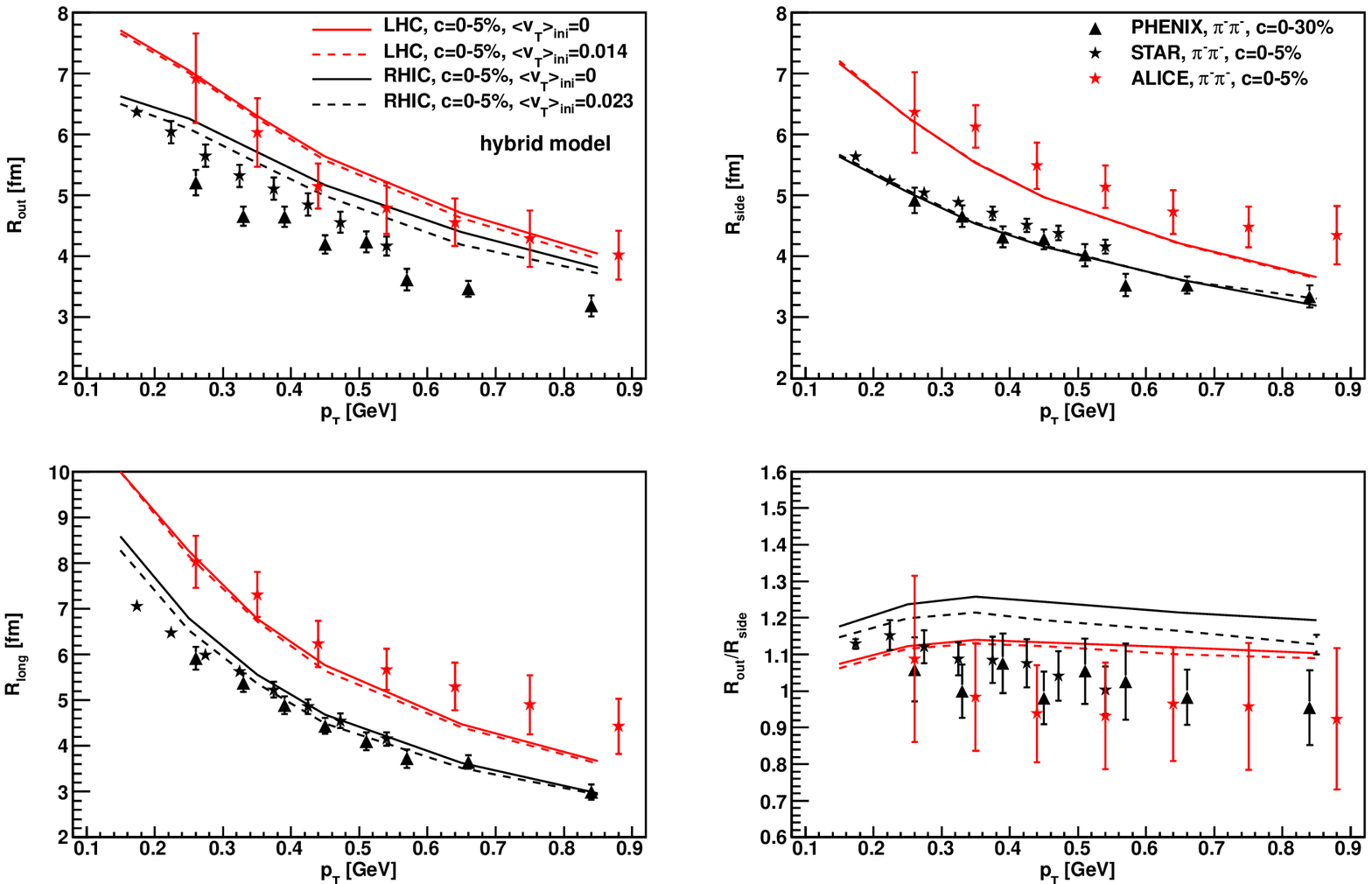}
 \caption{(Color online) $\pi^-$,$K^-$ and proton spectra and elliptic flow of all charged hadrons at midrapidity for 2.76 TeV LHC energy and different centrality classes (top) and HBT radii of $\pi^-$ pairs for most central events, calculated in the ``hybrid'' model and compared to experimental data from ALICE \cite{alice-spectra,alicev2,aliceHBT}. Solid lines correspond to IC without transverse flow, while dashed  lines stand for small initial transverse flows, $\left\langle v_T\right\rangle$=0.014 -- 0.018. Corresponding HBT radii for the top RHIC energy are shown for comparison purposes for the basic case without initial transverse flow. Note that calculations for $c=10-20$ \%  are not provided, but the $v_2$-coefficient for $c=0-5$ \% is presented instead as prediction.}\label{fig-lhc-iflow}
\end{figure*}

From the comparison with the LHC experimental data in Fig.\ref{fig-lhc-iflow} one can see some improvement of the behavior for all the observables, when small initial flow is included with the average value being $\left\langle v_\text{T,ini}\right\rangle$=0.014 -- 0.018. Namely, the effective temperature of the single-particle transverse spectra at large $p_T$ increases, while the $v_2$-coefficient decreases and $R_\text{out}/R_\text{side}$ ratio becomes lower.  

These ``viscous-like'' corrections bring the hope that the good description of the RHIC and LHC experiments can be reached within the HKM based on viscous hydrodynamics. This will be the next step in our study.

\section{Conclusions}
We present a consistent description of the transverse momentum spectra for the most abundant hadrons ($\pi$, $K$, $p$), elliptic flow coefficients, and interferometry radii for central and noncentral collisions at the top RHIC and LHC energies in the hybrid hydrokinetic model (hHKM). The latter provides the correct matching of decaying hadron matter evolution with hadronic UrQMD cascade at isochronic (or space-like in general) hypersurface. The results are compared with ones obtained at the different matching procedures in hybrid models with the same initial conditions. It is found that the matching of  hydrodynamics with cascade at the chemical freeze-out hypersurface gives the close results to hHKM, while the results are essentially different from both of these models and from experimental data when the matching of thermally and chemically equilibrated hydrodynamic systems with UrQMD happens at the isochronic hypersurface.  

The initial conditions allowing simultaneous description of the soft observables in ultrarelativistic A+A collisions are associated with the MC-Glauber initial entropy density distribution with mixed contributions from wounded nucleon density and density of binary scatterings. One can see the satisfactory description of the RHIC and LHC data within perfect hydrodynamics for the QGP phase, except for some overestimation of $v_2$ coefficients and $R_{out}/R_{side}$ ratios caused, probably, by neglect of the shear viscosity. As it is known the latter increases the final transverse flow and damps the elliptic flows. Qualitatively, similar effects can be reached by artificial inclusion of small but non-zero  initial transverse flow. We demonstrate that such an imitation of the shear viscosity indeed improves the $v_2$ coefficients and $R_{out}/R_{side}$ ratios. 

The hadron yields and particle number ratios for most abundant hadrons 
measured by ALICE are also well described and explained in the model. The contributions of different processes at the hadronic cascade stage to the final hadron yields are found.

Note that the most of the parameters fixed from agreement with RHIC data are applied also for the description of 2.76 TeV Pb+Pb LHC collisions with different centrality classes.
In fact, the only changes are the normalization of initial entropy density distribution, contribution from binary collisions  and baryonic chemical potentials at chemical freeze-out, which become zero. Then  pion, kaon and (anti) proton spectra, $p_T$-differential elliptic flow coefficients, and HBT radii are reproduced well also at LHC. It is the evidence of the similarity of the soft physics at the RHIC and LHC energies.
 
As the next step it is planned to include some missing features of the hydrokinetics into future versions of the model, namely the prethermal dynamics and viscosity.

\section{Acknowledgment}
The authors are grateful to S.V. Akkelin for discussions. The research was carried out within the scope of the EUREA: European Ultra Relativistic
Energies Agreement (European Research Group: ``Heavy ions at ultrarelativistic energies''), and is supported by the National Academy of Sciences of Ukraine (Agreement-2013) and by the State fund for fundamental researches of Ukraine (Agreement-2013). Iu.K. acknowledges the financial support by the ExtreMe Matter Institute EMMI and Hessian LOEWE initiative.

\end{document}